\documentclass[aip,jcp,amsmath,amsfonts,amssymb,twocolumn,reprint]{revtex4-1}
\usepackage{graphicx}
\usepackage{bm}
\usepackage{color}
\usepackage[colorlinks=true,linkcolor=blue,citecolor=blue,breaklinks]{hyperref}

\begin{document}

\title{Merging symmetry projection methods with coupled cluster theory: 
\\
Lessons from the Lipkin model Hamiltonian}

\author{Jacob M. Wahlen-Strothman}
\affiliation{Department of Physics and Astronomy, Rice University, Houston, TX, 77005, USA}

\author{Thomas M. Henderson}
\affiliation{Department of Physics and Astronomy, Rice University, Houston, TX, 77005, USA}
\affiliation{Department of Chemistry, Rice University, Houston, TX, 77005, USA}

\author{Matthew R. Hermes}
\affiliation{Department of Chemistry, Rice University, Houston, TX, 77005, USA}

\author{Matthias Degroote}
\affiliation{Department of Chemistry, Rice University, Houston, TX, 77005, USA}

\author{Yiheng Qiu}
\affiliation{Department of Chemistry, Rice University, Houston, TX, 77005, USA}

\author{Jinmo Zhao}
\affiliation{Department of Chemistry, Rice University, Houston, TX, 77005, USA}

\author{Jorge Dukelsky}
\affiliation{Instituto de Estructura de la Materia, CSIC, Serrano 123, E-28006 Madrid, Spain}

\author{Gustavo E. Scuseria}
\affiliation{Department of Physics and Astronomy, Rice University, Houston, TX, 77005, USA}
\affiliation{Department of Chemistry, Rice University, Houston, TX, 77005, USA}

\date{\today}

\begin{abstract}
Coupled cluster and symmetry projected Hartree-Fock are two central paradigms in electronic structure theory. However, they are very different. Single reference coupled cluster is highly successful for treating weakly correlated systems, but fails under strong correlation unless one sacrifices good quantum numbers and works with broken-symmetry wave functions, which is unphysical for finite systems. Symmetry projection is effective for the treatment of strong correlation at the mean-field level through multireference non-orthogonal configuration interaction wavefunctions, but unlike coupled cluster, it is neither size extensive nor ideal for treating dynamic correlation. We here examine different scenarios for merging these two dissimilar theories. We carry out this exercise over the integrable Lipkin model Hamiltonian, which despite its simplicity, encompasses non-trivial physics for degenerate systems and can be solved via diagonalization for a very large number of particles. We show how symmetry projection and coupled cluster doubles individually fail in different correlation limits, whereas models that merge these two theories are highly successful over the entire phase diagram. Despite the simplicity of the Lipkin Hamiltonian, the lessons learned in this work will be useful for building an ab initio symmetry projected coupled cluster theory that we expect to be accurate in the weakly and strongly correlated limits, as well as the recoupling regime.
\end{abstract}

\maketitle

\section{Introduction}
Single reference coupled cluster (CC) theory\cite{Bartlett07} has long been a central paradigm in quantum chemistry calculations due to its combination of high accuracy and reasonable polynomial computational cost which enables it to be applied to fairly sizable systems.

However, coupled cluster builds upon some underlying mean-field description of the wave function, typically Hartree-Fock (HF).  When the underlying mean field is qualitatively correct, so that the system is weakly correlated, coupled cluster theory is very effective.  Unfortunately, strong correlation is ubiquitous in materials and condensed matter, where it produce novel physics such as superconductivity, and is also common in chemistry.  The strongly correlated regime is poorly described in a single particle approximation, and consequently single-reference coupled cluster theory fails catastrophically.\cite{Bulik15} 

Not all hope is lost, however.  Strong correlation is intimately associated with degeneracies or near-degeneracies of the ground state, and in most cases these degeneracies are associated with some symmetry of the Hamiltonian.  Whereas symmetry restricted Hartree-Fock (RHF) theory is qualitatively incorrect in such strongly correlated problems, this failure is flagged by spontaneous symmetry breaking in the mean-field treatment.  By breaking symmetry, the mean-field description of the problem can mimic some of the effects of strong correlation at the cost of good quantum numbers, leading to the so-called ``symmetry dilemma'' in which one must choose between good quantum numbers on the one hand and a lower Hartree-Fock energy on the other.  Even when coupled cluster built atop the symmetry-adapted RHF fails, coupled cluster built on the broken-symmetry Hartree-Fock is often energetically reasonable.  Ultimately, however, symmetry-broken coupled cluster is not a panacea, as the broken symmetry compromises the quality of properties other than the energy, and symmetry-broken methods suffer from inaccuracies in the recoupling regime where both weak and strong correlation are important.  

What is needed is some way to take advantage of the strengths of broken-symmetry mean-field treatments without inheriting their weaknesses.  A conceptually straightforward way in which to do so is provided by symmetry projection.\cite{Jimenez12}  In this approach, a broken symmetry reference determinant is optimized in the presence of symmetry projection operators. The resulting wavefunction is a multireference non-orthogonal configuration interaction and is effective for treating strong correlation in small to medium size systems.\cite{JimenezHoyos2013,Samanta2012,Rivero2013,Rivero2013b}  However, symmetry projection of mean field wavefunctions is limited in its effectiveness for treating the dynamic correlations that coupled cluster so efficiently describes, and produces no improvement over broken-symmetry mean field in the thermodynamic limit.\cite{Jimenez12}

In this paper, we employ the Lipkin Hamiltonian,\cite{Lipkin65,Meshkov65,Glick65,Robinson89}  which we discuss in Sec. \ref{Sec:LipkinModel}, as a testbed for several scenarios where we merge coupled cluster and projected Hartree-Fock (PHF) theories.  In this Hamiltonian, the symmetry of interest for our purposes is known as parity, and we discuss parity projection and different ways to combine coupled cluster and parity projection in Sec. \ref{Sec:ParityProjection}.  As coupled cluster and PHF describe different types of correlation, we explore the merger's potential to produce a method that is accurate for all types of correlation while also maintaining good quantum numbers in the wavefunction.  We envision two basic scenarios.  One possibility is to work in the symmetry adapted basis but to introduce symmetry projection as a second set of particle-hole excitations, while the other is to projectively restore the symmetry of a broken-symmetry coupled cluster calculation.  The former is a model we have recently explored in an interpolation between coupled cluster and projected BCS theories,\cite{Degroote16} but rather than interpolating between the two approaches, here we wish to combine them following ideas we have recently outlined.\cite{Qiu16}  The latter is an excursion into symmetry-projected unrestricted coupled cluster theory, a model which has received some theoretical attention but which does not yet appear to have been implemented.\cite{Duguet15}  Section \ref{Sec:Results} shows results for each case.  We consider a variational treatment (\textit{i.e.} we obtain the energy and wave function parameters using the Rayleigh-Ritz variational principle) as well as similarity-transformed approaches which are closer in spirit to traditional coupled cluster theory.   We also present results for extended coupled cluster,\cite{Arponen1983b,Arponen1987,Arponen1987b,Piecuch1999,Fan2006,Cooper2010,Evangelista2011} the doubly similarity transformed theory of excitations and de-excitations that has received limited attention in quantum chemistry.  Generally speaking, the combination of similarity-transformation and symmetry projection already yields results in excellent agreement with the exact answer, and agreement is only improved by considering extended coupled cluster theory or using a variational approach.  We have hidden several mathematical results in the appendix so as not to clutter the main body of our work with extraneous detail, but the interested reader might find these results useful.

\section{The Lipkin Model
\label{Sec:LipkinModel}}
The Hamiltonian here referred to as the Lipkin model was proposed by Lipkin, Meshkov, and Glick as a simple yet non-trivial model of a closed shell nucleus with schematic monopole residual interactions. The simplicity of the model made it an excellent benchmark for testing many-body approximations, including time-dependent HF,\cite{Krieger74} extended coupled-cluster theory,\cite{Arponen83,Robinson89} time-dependent coupled cluster theory,\cite{Hoodbhoy79} self-consistent RPA,\cite{Dukelsky90}, symmetry projected HF,\cite{Robledo92} and generalized coupled cluster theory and the contracted Schr\"odinger equation.\cite{Mazziotti04}  In addition, the model is exactly solvable in terms of the bosonic Richardson ansatz,\cite{Ortiz2005,Lerma13} although the computational effort to solve the Richardson equations is higher than an exact diagonalization of the tridiagonal Hamiltonian matrix.

Loosely speaking, the Lipkin model for $n$ spinless fermions has two sets of single-particle levels, each of which is $n$-fold degenerate.  The Hamiltonian moves a pair of fermions from one set of levels to the other, but only ``vertically,'' \textit{i.e.} the Hamiltonian can only excite a fermion in the $i^\mathrm{th}$ lower level into the $i^\mathrm{th}$ upper level.  Importantly, only the \textit{number} of fermions in the upper level matters; which of the $n$ distinct upper levels are occupied does not.  The same is of course true for the lower level.  Accordingly, the number of parameters needed to express the exact wave function is \textit{linear} in the system size, not combinatorial.\cite{Lipkin65,Meshkov65,Glick65}  This simplification is of vital importance in the present context, as the methods we consider here have not yet been formulated into efficient generally applicable working theories.  Our goal is to generate highly accurate solutions with only a few parameters, which for problems with Hilbert spaces of combinatorial size would translate to methods with a polynomial number of parameters and hence a polynomial computational cost.

\begin{figure}[b]
\includegraphics[width=\columnwidth]{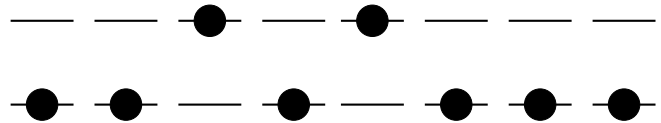}
\caption{Representative allowed state in the Lipkin Hamiltonian.  The upper and lower levels are both 8-fold degenerate, and occupied levels are denoted by filled dots. The Hamiltonian only creates vertical excitations or de-excitations, so this state would couple to the non-interacting ground state in which each of the lower levels is filled, but would not do so if the upper levels were occupied differently (\textit{i.e.} if either or both of the two particles in the upper level were directly above any of the particles in the lower level).
\label{Fig:Lipkin}}
\end{figure}

Mathematically, the Hamiltonian can be written compactly as
\begin{equation}
H = x \, J_z - \frac{1-x}{n} \, \left(J_+ \, J_+  +  J_- \, J_-\right)
\end{equation}
where $n$ is the number of particles and is equal to the degeneracy of each level, $x$ is an adjustable parameter, and the operators are
\begin{subequations}
\begin{align}
J_z &= \frac{1}{2} \, \sum_i \left(a^\dagger_{i,\uparrow} \, a_{i,\uparrow} - a^\dagger_{i,\downarrow} \, a_{i,\downarrow}\right),
\\
J_+ &= \sum_i a^\dagger_{i,\uparrow} \, a_{i,\downarrow},
\\
J_- &= \sum_i a^\dagger_{i,\downarrow} \, a_{i,\uparrow}.
\end{align}
\end{subequations}
We use $\downarrow$ and $\uparrow$ to respectively denote lower and upper single-particle levels.  As the notation is meant to indicate, the operators $J_+$, $J_-$, and $J_z$ satisfy SU(2) commutation relations:
\begin{subequations}
\begin{align}
[J_z,J_\pm] &= \pm J_\pm,
\\
[J_+,J_-] &= 2 \, J_z.
\end{align}
\end{subequations}
The Hamiltonian thus commutes with the operator
\begin{equation}
J^2 = \frac{J_+ \, J_- + J_- \, J_+}{2} + J_z \, J_z
\end{equation}
and its eigenstates can be labeled by the quantum number $j$.  The ground state of the Hamiltonian is contained in the subspace with $j = \frac{n}{2}$,\cite{Lipkin65,Meshkov65,Glick65} and this further simplifies the problem as we need only consider this block of the Hilbert space.  A representative state is shown in Fig.~\ref{Fig:Lipkin}.

The parameter $x$, taken to lie in [0,1], controls the relative importance of the interaction between the fermions; the problem is weakly correlated when $x$ is large and strongly correlated when $x$ is small.  In other words, the Hamiltonian is weakly correlated when the single-particle part of the Hamiltonian is the dominant contribution and is strongly correlated when the two-body part of the Hamiltonian is dominant.  The mean-field ground state for the weakly correlated case puts all the particles in the lower level, and $J_+$ then plays the role of a single excitation operator.  In other words, the full coupled cluster wave function could be written as 
\begin{equation}
|\Psi_{\mathrm{RCC}}\rangle = \mathrm{e}^{\sum t_i \, J_+^i} |0\rangle,
\end{equation}
where we emphasize that there is one and only one excitation amplitude $t_i$ per excitation level; $|0\rangle$ is the mean-field reference determinant
\begin{equation}
|0\rangle = \prod a_{i,\downarrow}^\dagger |-\rangle,
\end{equation}
and $|-\rangle$ is the physical vacuum.  This symmetry-adapted single determinant is the analog of the RHF determinant for the electronic Hamiltonian, and in a slight abuse of terminology we will refer to the symmetry-adapted determinant and coupled cluster as restricted Hartree-Fock and restricted coupled cluster (RCC).

In the strongly correlated limit, the lowest energy mean-field solution breaks a symmetry known as parity.  The parity of a state depends on the difference between the number of particles in the upper level and the number of particles in the lower level.  For systems with $n=4k$, parity is even if the number of particles in the upper level is even, and odd if the number of particles in the upper level is odd.  Mathematically, the parity operator is 
\begin{equation}
\Pi = \mathrm{e}^{\mathrm{i} \, \pi \, J_z}
\end{equation}
and has eigenvalues $\pm 1$.  Note that parity symmetry means that the symmetry-adapted coupled cluster has $t_{2k+1} = 0$; the exact ground state wave function consists only of even excitations.

We can represent the parity-broken mean-field state in terms of the single-particle operators $\alpha$ as
\begin{equation}
\begin{pmatrix} \alpha^\dagger_{i,\uparrow} \\ \alpha^\dagger_{i,\downarrow} \end{pmatrix}
= \frac{1}{\sqrt{1 + \kappa^2}} \, \begin{pmatrix} 1 & \kappa \\ -\kappa & 1 \end{pmatrix} \, \begin{pmatrix} a^\dagger_{i,\uparrow} \\ a^\dagger_{i,\downarrow} \end{pmatrix}
\label{Eqn:BSBasis}
\end{equation}
where recall that the operators $a$ refer to the symmetry-adapted basis (which is identical to the bare fermion basis for this problem).  The parity-broken ground state determinant is just
\begin{equation}
|\Phi\rangle = \prod_i \alpha_{i,\downarrow}^\dagger |-\rangle.
\end{equation}
Excited determinants in this broken-symmetry basis can be written by acting the operator
\begin{equation}
K_+ = \sum_i \alpha_{i,\uparrow}^\dagger \, \alpha_{i,\downarrow}
\end{equation}
on the broken-symmetry determinant $|\Phi\rangle$ in a way entirely analogous to the way excited determinants in the symmetry-adapted basis are created by the action of $J_+$ on the symmetry-adapted determinant $|0\rangle$.  Accordingly, broken-symmetry coupled cluster writes the wave function as
\begin{equation}
|\Psi_{\mathrm{UCC}}\rangle = \mathrm{e}^{\sum u_i \, K_+^i} |\Phi\rangle.
\label{Eqn:PsiUCC}
\end{equation}
The broken-symmetry determinant is the analog of UHF for the electronic Hamiltonian, so we will use unrestricted Hartree-Fock (UHF) and unrestricted coupled cluster (UCC) to refer to broken-symmetry mean-field and coupled cluster, while ``PHF'' will refer to the parity-projected mean-field, optimized in a variation after projection style.

\begin{figure}[t]
\includegraphics[width=\columnwidth]{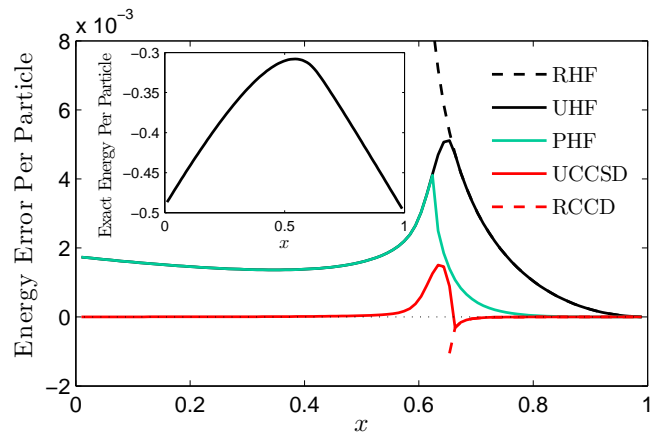}
\caption{Errors with respect to the exact result for the 50-particle Lipkin Hamiltonian.  By ``RHF'' and ``UHF'' we mean the Hartree-Fock solution which preserves or breaks parity symmetry; ``RCCD'' and ``UCCSD'' then refer to coupled cluster on top of the symmetry-adapted or symmetry-broken mean-field.  Note that single excitations on the symmetry-adapted reference are zero by symmetry.  We also show parity-projected broken-symmetry calculations, as ``PHF.''  The RCCD energy is complex for $x \lesssim 2/3$.
\label{Fig:ReferenceResults}}
\end{figure}

To see how this all plays out, Fig. \ref{Fig:ReferenceResults} shows total energies per particle as a function of $x$ for a variety of methods in the 50-particle Lipkin model (\textit{i.e.} $n = 50$, so the lower level and upper level are each 50-fold degenerate).  Note that RHF is only useful for $x \approx 1$, while UHF is qualitatively useful across the whole range of $x$.  The mean-field spontaneously breaks symmetry near $x = 2/3$, and indeed in the thermodynamic limit there is a second-order phase transition exactly at $x = 2/3$.  Projection improves upon the mean-field results significantly for large $x$ but has little impact for $x \lesssim 0.6$.  Both symmetry adapted coupled cluster (RCCD) and broken symmetry coupled cluster (UCCSD) are of exceptional quality in the weakly correlated limit, but where RCCD breaks down and delivers a complex energy as we move through the point of symmetry breaking, UCCSD is able to deliver virtually exact answers for all $x$ except for the intermediate coupling regime in the immediate vicinity of $x = 2/3$.  The price we pay is that while UCCSD is energetically accurate, it has strongly broken symmetry in the strongly correlated limit, which may have serious negative consequences for the evaluation of properties other than the energy.

Let us take a moment to emphasize why symmetry-adapted coupled cluster fails in the strongly correlated limit.  This is simply because, when truncated, the coupled cluster wave function places the wrong coefficients on relevant excited determinants.  If these coefficients are corrected, truncated coupled cluster-like schemes may be significantly improved.  Because symmetry-projected mean-field methods do describe strong correlations (and thus have more accurate coefficients of excited determinants), they constitute a promising means of providing this correction.\cite{Paldus1984,Piecuch1996,Degroote16,Qiu16}

\section{Parity Projection
\label{Sec:ParityProjection}}
Our goal in this work is to combine coupled cluster theory and symmetry projection.  The symmetry we wish to break and restore in this work is parity symmetry.  Although they are also symmetries of the Hamiltonian, we include neither number nor $J^2$ symmetry in this process.  Our experience suggests that we need only break and restore those symmetries which break spontaneously at the mean-field level.  This is simply because mean-field symmetry breaking is a manifestation of degeneracies at the Fermi level which are responsible for strong correlation.  While parity symmetry breaks spontaneously, neither  number nor $J^2$ are spontaneously broken.  Moreover, by working with a fixed number of particles in a fixed $j$ sector of Hilbert space, we can limit the number of parameters needed to describe the problem.

One can imagine expressing parity projection in many ways.  The simplest way is in terms of symmetry-adapted determinants, where as we have noted only even excitation levels contribute.  To that end, it will prove useful to write the broken symmetry determinant $|\Phi\rangle$ in terms of symmetry-adapted quantities.  We obtain
\begin{equation}
|\Phi\rangle = \frac{1}{\left(1 + \kappa^2\right)^{n/2}} \, \mathrm{e}^{\kappa \, J_+} |0\rangle.
\label{Eqn:DefPhi}
\end{equation}
Parity projection of $|\Phi\rangle$ simply eliminates the odd powers of $J_+$, so we have
\begin{subequations}
\begin{align}
P |\Phi\rangle 
 &= \frac{1}{2} \, \frac{1}{\left(1 + \kappa^2\right)^{n/2}} \, \left(\mathrm{e}^{\kappa \, J_+} + \mathrm{e}^{-\kappa \, J_+}\right) |0\rangle
\\
 &= \frac{1}{\left(1 + \kappa^2\right)^{n/2}} \, \cosh(\kappa \, J_+) |0\rangle
\end{align}
\end{subequations}
where $P$ projects parity symmetry.  This is the same sort of exercise as was carried out in our recent work on polynomial similarity transformations\cite{Degroote16} and on spin-projected Hartree-Fock.\cite{Qiu16}

Rather than referring to the PHF wave function by the action of particle-hole excitations acting on the symmetry-adapted determinant, we could write it in terms of excitations acting on the broken symmetry determinant instead.  Thus, we could write
\begin{equation}
P |\Phi\rangle = \mathcal{N} \, \left(1 + \mathrm{e}^{-2 \, \kappa \, J_+}\right) |\Phi\rangle
\end{equation}
where we have simply noted that $\mathrm{e}^{\kappa \, J_+} |0\rangle \sim |\Phi\rangle$ and $\mathcal{N}$ is a normalization constant.  The operator $J_+$ creates not only excitations but also de-excitations on the broken-symmetry determinant, as one can see by transforming it into the broken symmetry basis:
\begin{equation}
J_+ = \frac{K_+ - \kappa^2 \, K_- - 2 \, \kappa \, K_z}{1 + \kappa^2}
\end{equation}
where $K_-$ and $K_z$ are defined in analogy with $J_-$ and $J_z$:
\begin{subequations}
\begin{align}
K_- &= \sum_i \alpha^\dagger_{i,\downarrow} \, \alpha_{i,\uparrow},
\\
K_z &= \frac{1}{2} \, \sum_i \left(\alpha^\dagger_{i,\uparrow} \, \alpha_{i,\uparrow} - \alpha^\dagger_{i,\downarrow} \, \alpha_{i,\downarrow}\right).
\end{align}
\end{subequations}
One can then use a process known as disentanglement\cite{Gilmore05} to write the PHF wave function only in terms of excitations acting on the broken-symmetry determinant, as
\begin{equation}
P |\Phi\rangle = \frac{1}{1+S} \, \left(1 + S \, \mathrm{e}^{-2 \, \eta \, K_+}\right) |\Phi\rangle
\end{equation}
where $S$ and $\eta$ are
\begin{subequations}
\begin{align}
S &= \left(\frac{1 - \kappa^2}{1 + \kappa^2}\right)^n,
\\
\eta &= \frac{\kappa}{1 - \kappa^2}.
\end{align}
\end{subequations}

Once we wish to combine projected Hartree-Fock with coupled cluster theory, we have several possibilities.  Probably the simplest is to work in the symmetry-adapted picture.  Since only even excitations contribute, the symmetry preserving cluster operator is
\begin{equation}
T = \sum t_{2i} \, J_+^{2i}
\end{equation}
and the restricted coupled cluster form of the wave function is
\begin{equation}
|\mathrm{RCC}\rangle = \mathrm{e}^T |0\rangle.
\label{Def:PsiRCC}
\end{equation}
Because RCC preserves parity symmetry, nothing is gained by parity projection.  However, we can act the symetry-adapted cluster operator instead on a projected mean-field state to get what we will call the projected restricted coupled cluster (PRCC) form of the wave function:
\begin{equation}
|\mathrm{PRCC}\rangle = \mathrm{e}^T |\mathrm{PHF}\rangle = \mathrm{e}^T \, P \, |\Phi\rangle = P \, \mathrm{e}^T |\Phi\rangle.
\label{Def:PsiPRCC}
\end{equation}

One could also work in the broken symmetry basis from the beginning.  The broken-symmetry cluster operator is
\begin{equation}
U = \sum u_i \, K_+^i
\end{equation}
and an unrestricted coupled cluster wave function is generically
\begin{equation}
|\mathrm{UCC}\rangle = \mathrm{e}^U |\Phi\rangle.
\label{Def:PsiUCC}
\end{equation}
Adding a projection operator yields what we will refer to as projected unrestricted coupled cluster (PUCC):
\begin{equation}
|\mathrm{PUCC}\rangle = P \, \mathrm{e}^U |\Phi\rangle.
\label{Def:PsiPUCC}
\end{equation}
Analytic results in this case are more difficult to obtain, but thanks to the simplicity of the Lipkin Hamiltonian we can readily transform the broken-symmetry coupled cluster wave function back into the symmetry-adapted basis and delete those terms which break parity.

Of course in the limit where the cluster operator is complete, all of these approaches are entirely equivalent.  In practice, we will truncate the cluster operator to some low excitation level (\textit{e.g.} $T \approx t_2 \, J_+^2$), at which point we obtain different results depending on whether we work in the symmetry-adapted or broken-symmetry basis.  Generically, we follow the coupled cluster terminology in which $k$-fold excitations are created by $O_k$ for some operator $O$, and we note which excitation levels are included in the expansion.  Thus, for example, PRCC with double excitations (PRCCD) means
\begin{subequations}
\begin{align}
|\mathrm{PRCCD}\rangle &= \mathrm{e}^{T_2} \, P \, |\Phi\rangle,
\\
T_2 &= t_2 \, J_+^2,
\end{align}
\end{subequations}
while PUCC with single and double excitations (PUCCSD) means
\begin{subequations}
\begin{align}
|\mathrm{PUCCSD}\rangle &= P \, \mathrm{e}^{U_1 + U_2} |\Phi\rangle,
\\
U_1 &= u_1 \, K_+,
\\
U_2 &= u_2 \, K_+^2.
\end{align}
\end{subequations}

All projected methods naturally depend on the choice of the broken-symmetry reference $|\Phi\rangle$ or, in other words, upon the value of the symmetry breaking parameter $\kappa$.  We may use the value of $\kappa$ that minimizes the mean-field energy (denoted by $\kappa_\mathrm{UHF}$), or the value of $\kappa$ that minimizes the projected Hartree-Fock energy ($\kappa_\mathrm{PHF}$), or we may adjust $\kappa$ in the presence of the cluster operators.  We have suppressed the explicit $\kappa$-dependence in the equations to reduce notational clutter, and have explicitly indicated how $\kappa$ was obtained in all our results.

\section{Results
\label{Sec:Results}}
We are now in a position to explore the relative accuracy of our various approximations.  We face, however, an additional question in combining PHF and coupled cluster: how should we determine the energy and wave function amplitudes?  A variational determination is natural for PHF, while a similarity-transformed basis approach is natural from the perspective of coupled cluster theory.  While practicality in real systems probably requires the latter approach, in the Lipkin Hamiltonian both methods are feasible.  We will thus start by exploring the variational approach, just to see what the limitations of the wave function form might be.  We will continue to use the 50-particle model, as the number of sites is large enough to reduce finite size effects but not so large as to be computationally or numerically unwieldy.  For the most part, we limit ourselves to single and double excitations in our various cluster operators, so that we can explore approximations which might translate to practical calculations in more realistic Hamiltonians.

\subsection{Variational Treatment}
In our variational treatment, denoted by the addition of the prefix ``v'', we use a Hermitian expectation value for the energy, which we make simultaneously stationary with respect to all wave function parameters, including $\kappa$ (\textrm{i.e.} we obtain $|\Phi\rangle$ variationally in the presence of the exponential).  Thus, for example, the acronym vRCC implies that the wave function takes the restricted coupled cluster form of Eqn. \ref{Def:PsiRCC}, takes the energy as a Hermitian expectation value, and obtains the cluster amplitudes by making the energy stationary:
\begin{subequations}
\begin{align}
E_\mathrm{vRCC} &= \frac{\langle \mathrm{RCC} | H | \mathrm{RCC} \rangle}{\langle \mathrm{RCC} | \mathrm{RCC} \rangle},
\\
0 &= \frac{\partial E_\mathrm{vRCC}}{\partial t_i}.
\end{align}
\end{subequations}

\begin{figure}[t]
\includegraphics[width=\columnwidth]{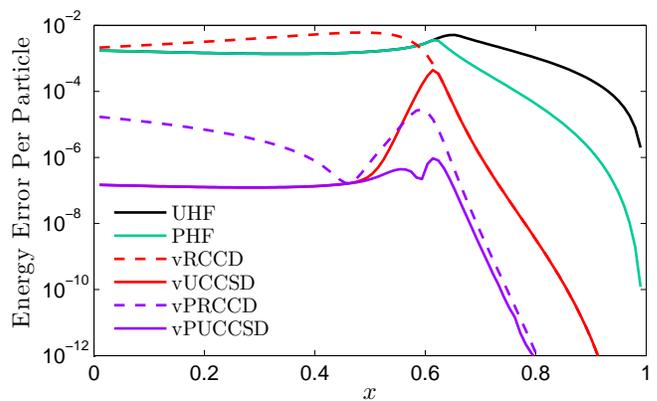}
\caption{Energy error of variational coupled cluster methods as a function of $x$ in the 50-particle Lipkin model. For small $x$, UHF coincides with PHF and vUCCSD coincides with vPUCCSD.
\label{Fig:VariationalResults}}
\end{figure}

Figure \ref{Fig:VariationalResults} shows errors per particle in the 50-particle Lipkin Hamiltonian.  Notice that PHF improves significantly upon UHF in the weakly correlated case, but has little effect past the phase transition near $x \sim 2/3$.  Variational coupled cluster is uniformly excellent, and is improved even further in the weak and intermediate coupling regime by variational projected coupled cluster.  This all serves to indicate that the wave function form is sufficiently flexible that highly accurate results can be obtained with only a few parameters in the wave function.  We should point out that for small $x$, the two parity sectors become nearly degenerate (and exactly so at $x = 0$), as a consequence of which PHF and UHF are almost degenerate, as are vUCCSD and vPUCCSD.

In general, we cannot expect to create an effective similarity-transformed method if the variational approach demonstrates that the wave function form we have chosen is inadequate.  Our results here show that a combination of PHF and coupled cluster is not doomed from the start.

\subsection{Similarity-Transformed Treatment}
While we have shown that projected coupled cluster wavefunctions can effectively approximate the exact wavefunction with only a few parameters, variational methods are not feasible for general systems without truncation.  In practice we will have to adopt a similarity-transformed approach which allows us to work with a short expansion of the projected CC wave function rather than requiring the entire thing.  Unfortunately there is no guarantee that what works in a variational context will work in the similarity-transformed context as well.  Indeed, where vRCCD is highly accurate for all $x$, RCCD itself has no (real) solution for strongly correlated systems.  Accordingly, we now turn our attentions to various similarity-transformed approximations, to find which inherit the capabilities of the variational approach and which do not.

Let us begin with projected restricted coupled cluster.  Our point of departure is the Schr\"odinger equation, written in the form
\begin{equation}
H \, P \, \mathrm{e}^{T} |\Phi\rangle = E \, P \, \mathrm{e}^{T} |\Phi\rangle.
\end{equation}
Because $P$ and $T$ commute, we could equivalently (and more conveniently) write it as
\begin{equation}
H \, \mathrm{e}^T |\mathrm{PHF}\rangle = E \, \mathrm{e}^T |\mathrm{PHF}\rangle.
\end{equation}
Premultiplying by $\mathrm{e}^{-T}$ defines the similarity-transformed Hamiltonian $\bar{H}$, and we have
\begin{equation}
\mathrm{e}^{-T} \, H \, \mathrm{e}^{T} |\mathrm{PHF}\rangle = \bar{H} |\mathrm{PHF} \rangle = E |\mathrm{PHF}\rangle.
\end{equation}
In other words, the PHF state is a right-hand eigenstate of $\bar{H}$ with the exact ground state eigenvalue.  We will also need the left-hand eigenstate of $\bar{H}$, which we write as
\begin{subequations}
\begin{align}
\langle \mathrm{PHF} | (1 + Z) \bar{H} &= E \, \langle \mathrm{PHF} | (1 + Z),
\\
Z  &= \sum Z_{2k},
\\
Z_{2k} &= z_{2k} \, J_-^{2k}.
\end{align}
\end{subequations}
With these ingredients in hand, we can write the projected restricted coupled cluster approach as
\begin{subequations}
\begin{align}
E_\mathrm{PRCC} &= \frac{\langle \mathrm{PHF} | (1 + Z) \, \mathrm{e}^{-T} \, H \, \mathrm{e}^T |\mathrm{PHF}\rangle}{\langle \mathrm{PHF} | 1 + Z |\mathrm{PHF}\rangle},
\\
0 &= \frac{\partial E_\mathrm{PRCC}}{\partial t_k},
\\
0 &= \frac{\partial E_\mathrm{PRCC}}{\partial z_k}.
\end{align}
\end{subequations}
Note that we do not need the $Z$ amplitudes to evaluate the energy.  We introduce them here to demonstrate how the linear response would be calculated.  Limiting $T$ and $Z$ to double excitations and de-excitations gives projected restricted coupled cluster doubles.

We still need to specify how we obtain the broken-symmetry determinant used to construct the projected mean-field state, which we recall is
\begin{equation}
|\mathrm{PHF}\rangle = P |\Phi\rangle \sim P \, \mathrm{e}^{\kappa J_+} |0\rangle;
\end{equation}
in other words, we must provide some way of calculating $\kappa$.  Frustratingly, taking the value of $\kappa$ that minimizes the mean-field energy ($\kappa_\mathrm{UHF}$) or the projected mean-field energy ($\kappa_\mathrm{PHF}$) proved inadequate; the equations for $t_2$ and $z_2$ frequently failed to converge and when they did converge we obtained unacceptably poor results.  Nor did making the energy stationary with respect to $\kappa$ prove to be particularly fruitful.  In the end, we selected a Brueckner-style approach in which we write
\begin{equation}
\langle \Phi| K_- \, \left(\bar{H} - E_\mathrm{PRCCD}\right) |\mathrm{PHF}\rangle = 0
\end{equation}
where $K_-$ is defined using the same value of $\kappa$ as was used to define $|\Phi\rangle$.  The value of $\kappa$ selected by this approach will be labeled as $\kappa_B$.

We can follow a basically similar line of reasoning to obtain projected unrestricted coupled cluster.  Here, we write the energy as
\begin{equation}
E_\mathrm{PUCC} = \frac{\langle \Phi| (1 + W) \, \mathrm{e}^{-U} \, H \, P \, \mathrm{e}^{U} |\Phi\rangle}{\langle \Phi| (1 + W) \, \mathrm{e}^{-U} \, P \, \mathrm{e}^{U} |\Phi\rangle},
\end{equation}
where we have introduced the broken symmetry de-excitation operator
\begin{subequations}
\begin{align}
W &= \sum W_k,
\\
W_k &= w_k \, K_-^k,.
\end{align}
\end{subequations}
Again we solve for the amplitudes $u_k$ and $w_k$ by requiring the energy to be stationary.  Truncating $U$ and $W$ to single and double excitations and de-excitations defines projected unrestricted coupled cluster singles and doubles.  As in the restricted case, we find that a Brueckner-style approach works best, and a Brueckner-style optimization of $\kappa$ just means finding the reference determinant $|\Phi\rangle$ such that the equations for $u_1$ are satisfied at $u_1 = 0$:
\begin{equation}
\langle \Phi| K_- \, \mathrm{e}^{- U_2} \, \left(H - E_\mathrm{PUCCD}\right) \, P \, \mathrm{e}^{ U_2} |\Phi \rangle = 0.
\end{equation}
Again, we will use $\kappa_B$ to define this Brueckner-style solution for $\kappa$.

\begin{figure}[t]
\includegraphics[width=\columnwidth]{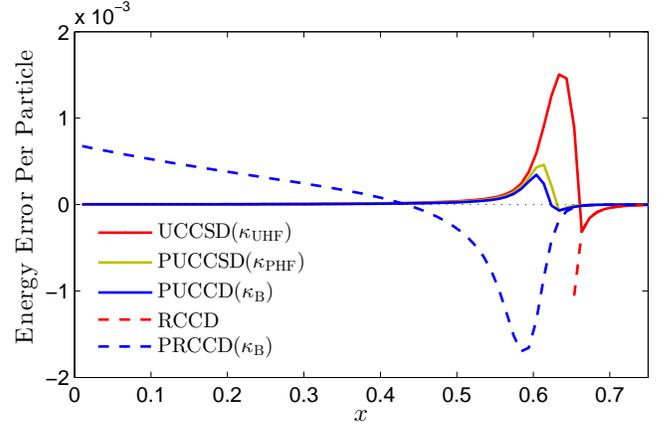}
\caption{Projected coupled cluster energies in the 50-particle Lipkin Hamiltonian.
\label{Fig:PCC}}
\end{figure}

Results from these various ideas are shown in Fig. \ref{Fig:PCC}.  It is clear that whereas RCCD fails disastrously except for $x \gtrsim 2/3$, PRCCD is at least qualitatively reasonable everywhere.  Agreement with the exact result is far from perfect, however.  The methods based on broken-symmetry coupled cluster fare better.  The broken symmetry coupled cluster (UCCSD) is already good except near the symmetry-breaking point.  Adding the projection operator improves the results further, regardless of whether we use the Brueckner procedure to define the broken-symmetry reference or we take the broken-symmetry reference from PHF (denoted by PUCCSD($\kappa_\mathrm{PHF}$) on the plot).

It should be noted that the PUCC formalism reduces to the PRCC approach in a natural way.  Indeed, if we begin with the PUCCSD equations and make the replacements $U \to T_2$ and $W \to Z_2$, we obtain PRCCD.  To see this, we use the fact the projection operator $P$ commutes with $T$ and $Z$ and is Hermitian and idempotent ($P = P^\dagger = P^2$).  In general, of course, the projection operator does not commute with $U$ or $W$, and the PUCC formalism, while more accurate, is also more cumbersome.

While we have noted the formal similarity between PRCC and PUCC, it is perhaps more interesting to consider their differences.  This we can do by expressing everything in terms of symmetry-adapted operators $J_+$, $J_-$, and $J_z$.  At the simplest level of theory (one with only double excitations) we could write the PRCCD wave function as
\begin{equation}
|\mathrm{PRCCD}\rangle = P \, \mathrm{e}^{t_2 \, J_+^2} \, |\Phi\rangle
\end{equation}
while the PUCCD wave function becomes
\begin{equation}
|\mathrm{PUCCD}\rangle = P \, \mathrm{e}^{\tilde{t}_2 \, \left(J_+ - \kappa^2 \, J_- + 2 \, \kappa \, J_z\right)^2} \, |\Phi\rangle
\end{equation}
where we have expressed $K_+$ as a linear combination of $J_+$, $J_-$, and $J_z$, as
\begin{equation}
K_+ = \frac{J_+ - \kappa^2 \, J_- + 2 \, \kappa \, J_z}{1 + \kappa^2}
\end{equation}
and have defined
\begin{equation}
\tilde{t}_2 = \frac{u_2}{(1 + \kappa^2)^2}.
\end{equation}
In other words, with operators expressed in the symmetry-adapted basis, PUCC contains both excitation and de-excitation operators in the exponential while PRCC contains only excitation operators.  That PUCC delivers better results than does PRCC suggests that symmetry-adapted de-excitation operators may be important.  To this end, we now turn our attention to an extended coupled cluster variant of the theory.

\subsection{Extended Coupled Cluster}
While traditional coupled cluster theory does not require an optimized left-hand state to extract the energy, making nontrivial use of the left-hand state can significantly improve the results of coupled-cluster-style calculations.\cite{Arponen1983b,Byrd2002,Piecuch2005,Qiu16}  To assess the impact of this effect here, we generalize our similarity-transformed methods to extended coupled cluster,\cite{Arponen1983b,Arponen1987,Arponen1987b,Piecuch1999} which includes a second similarity transformation, this time constructed with de-excitation operators.  

Put briefly, restricted and unrestricted extended coupled cluster (RECC and UECC) both define the energy by a double similarity transformation and make the energy stationary with respect to the parameters of the wave function.  Thus, we have
\begin{subequations}
\begin{align}
E_\mathrm{RECC} &= \langle 0| \mathrm{e}^Z \, \mathrm{e}^{-T} \, H \, \mathrm{e}^T \, \mathrm{e}^{-Z} |0\rangle,
\\
E_\mathrm{UECC} &= \langle \Phi | \mathrm{e}^W \, \mathrm{e}^{-U} \, H \, \mathrm{e}^U \, \mathrm{e}^{-W} |\Phi\rangle.
\end{align}
\end{subequations}
We can include projection operators in an essentially obvious way to define projected RECC and UECC (denoted as PRECC and PUECC):
\begin{subequations}
\begin{align}
E_\mathrm{PRECC} &= \frac{\langle \mathrm{PHF} | \mathrm{e}^{Z} \, \mathrm{e}^{-T} \, H \, \mathrm{e}^T \, \mathrm{e}^{-Z} |\mathrm{PHF}\rangle}{\langle \mathrm{PHF} | \mathrm{PHF}\rangle},
\\
E_\mathrm{PUECC} &= \frac{\langle \Phi| \mathrm{e}^{W} \, \mathrm{e}^{-U} \, H \, P \, \mathrm{e}^{U} \, \mathrm{e}^{-W} |\Phi\rangle}{\langle \Phi | \mathrm{e}^W \, \mathrm{e}^{-U} \, P \, \mathrm{e}^U \, \mathrm{e}^{-W} | \Phi\rangle}.
\end{align}
\end{subequations}
As in extended coupled cluster, we will solve for the coefficients in $T$, $U$, $W$, and $Z$ by making the energy stationary.  Truncating $T$ and $Z$ at double excitations and de-excitations gives projected restricted extended coupled cluster doubles (PRECCD) and truncating $U$ and $W$ to single and double excitations and de-excitations gives projected unrestricted extended coupled cluster singles and doubles (PUECCSD).  Note that in the unrestricted case we could take advantage of the fact that $W|\Phi\rangle=0$ to simplify the expression slightly; this is not possible for the restricted case, because
\begin{equation}
Z |\mathrm{PHF}\rangle = Z \, P \, |\Phi\rangle = P \, Z | \Phi\rangle
\end{equation}
and $Z$ is a de-excitation operator when acting on $|0\rangle$, not on $|\Phi\rangle$.

In both the restricted and unrestricted case, we find that orbital optimization is well-behaved (and for the restricted case, necessary), and orbital-optimized methods accordingly define the broken symmetry reference determinant by solving
\begin{equation}
\frac{\partial E}{\partial \kappa} = 0.
\end{equation}
For the restricted case, this amounts to minimizing the expectation value of the doubly similarity-transformed Hamiltonian $\mathrm{e}^Z \, \bar{H} \, \mathrm{e}^{-Z}$ over projected mean-field states.  The value of $\kappa$ found by making the energy stationary in this manner will be referred to as $\kappa_\mathrm{opt}$.  Note that orbital optimization is not readily compatible with the inclusion of single excitations.

\begin{figure}[t]
\includegraphics[width=\columnwidth]{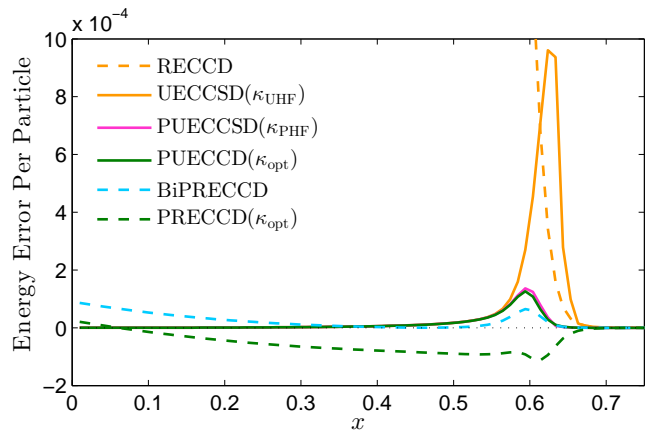}
\caption{Projected extended coupled cluster energies in the 50-particle Lipkin Hamiltonian.
\label{Fig:PECC}}
\end{figure}

In Fig. \ref{Fig:PECC}, we show calculated energies for PECC, and compare to traditional ECC.  While RECCD is highly inaccurate for intermediate to strong coupling, adding projection and orbital optimization dramatically improved the results.  Further improvement was possible by using different reference determinants to the left and the right, in a biorthogonal projected extended coupled cluster (BiPRECC), in which we write
\begin{equation}
E_\mathrm{BiPRECC} = \frac{\langle \Phi'| \mathrm{e}^Z \, \bar{H} \, \mathrm{e}^{-Z} \, P \, |\Phi\rangle}{\langle \Phi'| P | \Phi\rangle}.
\end{equation}
In this case, we make the energy stationary separately with respect to the rotations $\kappa$ defining $|\Phi\rangle$ and $\kappa'$ defining $|\Phi'\rangle$.

It is clear that the largest error for the symmetry-adapted  results is in the strong correlation regime.  In this limit extended unrestricted coupled cluster is very accurate, which is to be expected as the traditional unrestricted coupled cluster was already nearly exact.  The method is less accurate in the intermediate coupling regime, but can be significantly improved by adding projection.  Thus, PUECCSD with $\kappa$ taken from PHF is already highly accurate everywhere, and little improvement is to be found by orbital optimization,  However, even the best unrestricted model has greater error at intermediate coupling than we found simply by combining orbital optimization with in a biorthogonal approach.  We attempted to further improve the unrestricted results with a biorthogonal ansatz, but this made little difference as the left and right determinants converged to nearly identical states.

Finally, we note that the improvement of PRECC over PRCC suggests that it may be beneficial to use a more sophisticated left-hand state for PRCC than the simple linear ansatz we have chosen.  We do not explore that possibility here, but a sort of scheme intermediate between PRCC and PRECC in which we write
\begin{equation}
E = \frac{\langle \mathrm{PHF} | \mathrm{e}^{Z} \, \mathrm{e}^{-T} \, H \, \mathrm{e}^T |\mathrm{PHF}\rangle}{\langle \mathrm{PHF} | \mathrm{e}^Z |\mathrm{PHF}\rangle}
\end{equation}
may be worth pursuing.

\subsection{Higher Excitations}
Thus far, we have limited ourselves only to single and double excitations in the cluster operator.  Although even these relatively low-order projected coupled cluster methods are already fairly accurate, we would like to see how rapidly projected coupled cluster converges toward the exact answer as we include higher excitations.  In this section we return to the variational methodology, which is numerically more robust than are the similarity-transformed approaches.  As we show below, while the variational methods are also marginally more accurate, the difference is not too striking; accordingly, we would expect the two versions of the theory to show similar qualitative dependence on the truncation of the cluster operator.

\begin{figure}[b]
\includegraphics[width=\columnwidth]{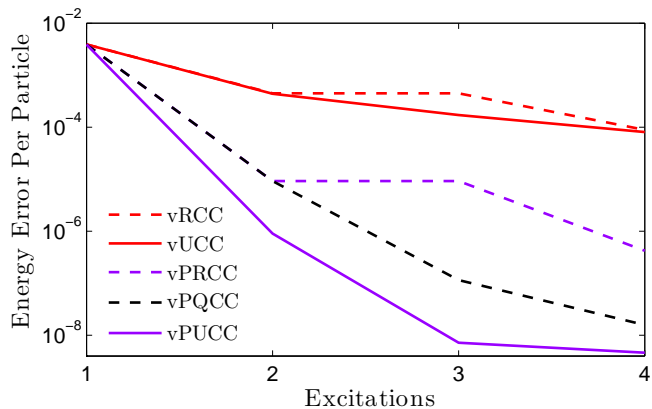}
\caption{Energy error in various coupled cluster models for the 50-particle Lipkin model at $x = 0.62$ as a function of highest excitation level included.
\label{Fig:HigherExcitations}}
\end{figure}

Figure \ref{Fig:HigherExcitations} therefore shows how various variational coupled cluster methods behave as the cluster operator is made more and more complete.  We look near $x = 2/3$, as it is in this recoupling regime that we expect to see largest errors.  We emphasize that, as with our previous variational results, we make $\kappa$ an additional variational parameter in vPRCC and vPUCC; we will introduce vPQCC momentarily.

First, it is clear that projected coupled cluster converges much more rapidly than does traditional coupled cluster.  By far the best approach seems to be projecting the broken-symmetry coupled cluster wave function (PUCC), which is to be expected as this is the most flexible ansatz.  Omitting the projection operator yields results which converge toward the exact answer much slower.  

The restricted methods of course show improvement only at even excitation levels, the odd excitations vanishing, as they should, as a consequence of parity symmetry.  We can, however, considerably improve upon PRCC in what we have called projected quasirestricted coupled cluster (labeled in Fig. \ref{Fig:HigherExcitations} as vPQCC).  The idea is very simple.  The broken symmetry determinant $|\Phi\rangle$ used in PRCC is created by the exponential of a single excitation operator, so we could write
\begin{equation}
|\mathrm{PRCC}\rangle = P \, \mathrm{e}^{T + Q_1} |0\rangle
\end{equation}
where $Q_1 = \kappa \, J_+$ creates $|\Phi\rangle$.  This operator $Q_1$ is the first in an entire family of operators 
\begin{subequations}
\begin{align}
Q_{2k+1} &= q_{2k+1} \, J_+^{2k+1},
\\
q_1 &= \kappa.
\end{align}
\end{subequations}
One could use these operators to write what we have chosen to call the quasirestricted coupled cluster wave function:
\begin{subequations}
\begin{align}
|\mathrm{QCC}\rangle &= \mathrm{e}^{T + Q} |0\rangle.
\\
Q &= \sum Q_{2k+1}.
\end{align}
\end{subequations}
Because $Q$ breaks symmetry, in the absence of a projection operator the amplitudes defining $Q$ vanish, and we might as well exclude $Q$ entirely.  However, when we add a projection operator to obtain projected quasirestricted coupled cluster, $Q$ contributes to the wave function:
\begin{equation}
|\mathrm{PQCC}\rangle = P \, \mathrm{e}^{T + Q}|0\rangle = \mathrm{e}^T \, \cosh(Q) |0\rangle.
\end{equation}
Figure \ref{Fig:HigherExcitations} shows that there is much to gain by including $Q_3$.  Note that when $Q$ is truncated at single excitations, PQCC reduces to PRCC.

\begin{figure}[t]
\includegraphics[width=0.49\textwidth]{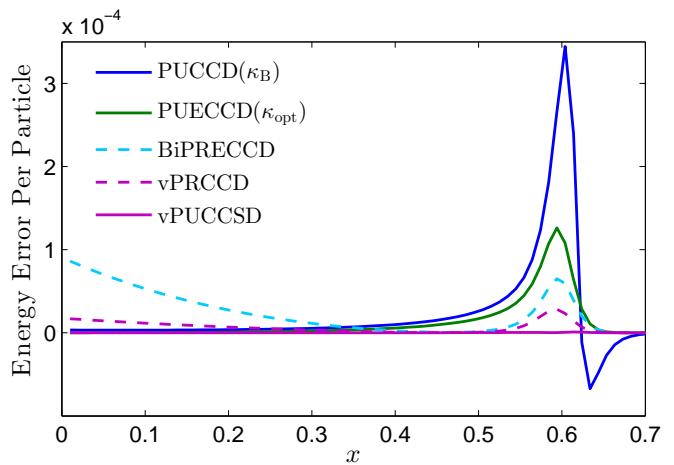}
\caption{A comparison of variational and similarity transformed energies for the Lipkin model with 50 particles.
\label{Fig:FinalComparison}}
\end{figure}

\subsection{Comparing Variational and Similarity-Transformed Approaches}
While similarity-transform methods are more feasible to implement than are variational forms, it is important to see how much quality they sacrifice.  This can be seen from Fig. \ref{Fig:FinalComparison}, which compares selected variational and similarity-transformation-based approaches.  While the variational results overall are of slightly better quality, the similarity-transform-based calculations are not significantly less accurate.  The extended coupled cluster methods in particular are in very strong agreement with the variational energies; for example, vPUCCD and PUECCD differ by at most $\mathcal{O}(10^{-9})$, and BiPRECCD is of accuracy comparable to the other methods everywhere but in the strongly correlated limit.  By far the most accurate approach is vPUCCSD, for which we do not have a good analog in our similarity-transformation-based schemes because we cannot in general simultaneously include single excitations and orbital optimization without introducing serious numerical difficulties.

\section{Conclusions}
In the search for a method capable of accurately treating weak and strong correlation both individually and together in the recoupling region, symmetry projected coupled cluster is a promising candidate. We have shown that it produces highly accurate results for the Lipkin Hamiltonian across all interaction strengths in both the variational and similarity transformed methods.  Our results here are a good indicator that these methods will produce significant improvements over traditional coupled cluster theory.

We should emphasize that much additional work is required to apply this methodology to other symmetries in other systems.  The fully variational treatment is numerically easiest but is computationally unfeasible for realistic Hamiltonians, and similarity-transform-based methods are therefore required.  Probably the simplest such approach is generalizing PRCC, but even there we are faced with the formidable (but manageable) task of evaluating the expectation value of $\bar{H}$ with respect to a PHF wave function.  Using the simple structure of PHF density matrices\cite{Jimenez12}, one could evaluate $\langle \mathrm{PHF} | \bar{H} |\mathrm{PHF}\rangle$ and $\langle \mathrm{PHF} | Z \, \bar{H} | \mathrm{PHF}\rangle$ with $\mathcal{O}(N^8)$ scaling at the PRCCSD level.  The scaling could be reduced  if the five- and six-body terms in $\bar{H}$ could be neglected, but it is not yet clear whether doing so yields sufficiently accurate results.  Nor is it clear how PRCC behaves as a function of system size; while PHF reduces to UHF in the thermodynamic limit, the limiting behavior of the coupled cluster correction to PHF is not yet obvious.

While significant work remains, none of it seems insurmountable, and preliminary investigations in other model Hamiltonians and a few small molecular examples suggests that the encouraging results we see here for the Lipkin Hamiltonian are not misleading.  A theory that can be relied on to produce accurate results for all types of correlation will be a significant step forward in the study and understanding of novel molecules and materials and a valuable tool for quantum mechanical research.

\begin{acknowledgments}
This work was supported by the U.S. Department of Energy, Office of Basic Energy Sciences, Computational and Theoretical Chemistry Program under Award No. DE-FG02-09ER16053. G.E.S. is a Welch Foundation Chair (No. C-0036).  Computational resources for this work were supported in part by the Big-Data Private-Cloud Research Cyberinfrastructure MRI-award funded by NSF under grant CNS-1338099 and by Rice University.  J.D. acknowledges support from the Spanish Ministry of Economy and Competitiveness and FEDER through Grant No. FIS2015-63770-P.
\end{acknowledgments}

\appendix
\section{Useful Formulae}
In this appendix, we collect several formulae and mathematical results which might be of use to anyone seeking to use the Lipkin Hamiltonian in their own work.  We include analytic expressions for the Hartree-Fock energy and the matrix elements of the full configuration interaction Hamiltonian.

We have noted that the Hamiltonian can be represented in terms of angular momentum operators $J_+$, $J_-$, and $J_z$, that $J^2$ is therefore a symmetry, and that the ground state has angular momentum $j = n/2$.  We can form a useful basis from the $J_z$ eigenstates with total angular momentum $j$.  The ground state at $x=1$ places all the particles in the lower level, which in this angular momentum language is the state with angular momentum $j$ and $J_z$ eigenvalue $-j$.  Excited states are created by applying $J_+$.  Labeling the state with $J_z = -j+m$ as $|m\rangle$, so that the ground state is $|0\rangle$, our basis is defined by
\begin{subequations}
\begin{align}
J_z |m\rangle &= \left(m - \frac{n}{2}\right) |m\rangle,
\\
J_+ |m\rangle &= \sqrt{(m+1)\, (n-m)} |m+1\rangle,
\\
J_- |m\rangle &= \sqrt{m \, (n-m+1)} |m-1\rangle.
\end{align}
\end{subequations}
The coefficients here account for the degeneracies in the energy levels and act as a convenient bookkeeping tool when evaluating expectation values.  They look rather unorthodox, but are recovered from the usual coefficients for ladder operators with $j=n/2$ and $j_z = -j+m$.

The full Hamiltonian is a $(n+1) \times (n+1) = (2j+1) \times (2j+1)$ matrix, with elements
\begin{subequations}
\begin{align}
\langle m | H | m\rangle
 &= x \langle m| J_z|m\rangle
\\
 &= x \, \left(m - \frac{n}{2}\right),
\nonumber
\\
\langle m | H | m+2\rangle
 &= \frac{x-1}{n} \, \langle m| J_- \, J_- |m+2\rangle
\\
 = \frac{x-1}{n}& \sqrt{m \, (m-1) \, (n-m+1) \, (n-m+2)},
\nonumber
\\
\langle m | H |m-2\rangle
 &= \frac{x-1}{n} \, \langle m| J_+ \, J_+ |m-2\rangle
\\
 = \frac{x-1}{n} & \sqrt{(m+2) \, (m+1) \, (n-m-1) \, (n-m)}.
\nonumber
\end{align}
\end{subequations}
Other entries vanish. Because we have a compact representation for the Hamiltonian, the exact energies of larger systems can easily be calculated.  The main limitation in size is potential numerical inaccuracy and overflow from the combinatorial elements of the Hamiltonian matrix.  The energy of the symmetry-adapted mean-field is just read off as $\langle 0|H|0\rangle$, and is thus
\begin{equation}
E_\mathrm{RHF} = -\frac{x \, n}{2}.
\end{equation}

For calculations involving a broken symmetry reference, it is useful to write the Hamiltonian in terms of the broken symmetry operators $K_+$, $K_-$, and $K_z$.  We obtain
\begin{align}
H 
 &= h_z \, K_z + h_\pm \, (K_++ \, K_-) + v_\pm \, (K_+ \, K_+ + K_-\, K_-)
\\
 &+v_z \, K_z \, K_z + v_\times \, K_+ \, K_- + v_{\pm z} \, (K_+ \, K_z + K_z \, K_-),
\nonumber
\end{align}
where
\begin{subequations}
\begin{align}
h_z &= \frac{x \, (1-\kappa^2)}{1+\kappa^2}-\frac{4 \, (1-x) \, \kappa^2}{n \, (1+\kappa^2)^2}
\\
h_\pm &= \frac{x \, \kappa}{1+\kappa^2}-\frac{2 \, \kappa \, (1-x) \, (\kappa^2-1)}{n \, (1+\kappa^2)^2}
\\
v_z &= -\frac{8 \, \kappa^2 \, (1-x)}{n \, (1+\kappa^2)^2}
\\
v_\pm &= -\frac{(1-x) \, (1+\kappa^4)}{n \, (1+\kappa^2)^2}
\\
v_\times  &= \frac{4 \, \kappa^2 \, (1-x)}{n \, (1+\kappa^2)^2}
\\ 
v_{\pm z} &= -\frac{4 \, \kappa \, (\kappa^2-1) \, (1-x)}{n \, (1+\kappa^2)^2}.
\end{align}
\end{subequations}
The Hamiltonian matrix elements can be expressed in terms of the broken symmetry states $|\tilde m\rangle$ created by the action of $K_+$ on the broken-symmetry reference state $|\tilde{0}\rangle = |\Phi\rangle$.  The UHF energy will simply be the element $\langle\tilde{0}|H|\tilde{0}\rangle = \langle \Phi|H|\Phi\rangle$, which we find to be
\begin{equation}
E_\mathrm{UHF}(\kappa) = E_\mathrm{RHF} + \frac{\kappa^2}{1 + \kappa^2} \, n \, x + \frac{2 \, \kappa^2}{(1 + \kappa^2)^2} \, (x-1) \, (n-1).
\end{equation}
Minimizing the energy with respect to $\kappa$ gives
\begin{equation}
\kappa_{0} =
\begin{cases}
0                                                                 & \quad \text{if } x \geq \frac{2n-2}{3n-2}
\\
\pm\sqrt{\frac{x(3n-2)-2n+2}{x(n-2)-2n+2}} & \quad \text{if } x < \frac{2n-2}{3n-2}
\\
\end{cases},
\end{equation}
so the symmetry-broken energy is identical to the symmetry-adapted energy for $x \ge (2n-2)/(3n-2)$ while for smaller $x$ we have
\begin{equation}
E_\mathrm{UHF}(\kappa_{0}) = \frac{(4-8n)(x-1)^2 +n^2(4+x(5x-8))}{8(n-1)(x-1)}.
\end{equation}

With more exertion, one can derive the energy of the PHF wave function as a function of $\kappa$, and we obtain
\begin{align}
%E_{PHF} = &E_{UHF} +\frac{1}{{(1+\kappa^2)^n + (1-\kappa^2)^n}}
%\\
%      \times &\Big[2 \, \kappa^2 \, (1-\kappa^2)^{n-2} \, (1+\kappa^2)^2 \, 
%\nonumber
%\\
%       &\times\big(n \, x \, (\kappa^4-1) + 4 \, \kappa^2 \, (n-1) \, (x-1) \big)\Big]
%\nonumber
E_\mathrm{PHF} &= E_\mathrm{UHF} - \frac{\left(1 - \kappa^2\right)^n}{\left(1 - \kappa^2\right)^n + \left(1 + \kappa^2\right)^n} \, \frac{2 \, \kappa^4}{\left(1 - \kappa^4\right)^2}
\\
  & \times \Big[2 \, \left(1 - 3 \, \kappa^2 + \kappa^4 - \kappa^6\right) \, \left(x-1\right) \, \left(n-1\right) 
\nonumber
\\
     & + \left(1 - \kappa^4\right) \, x \, n\Big]
\nonumber
\end{align}
We minimize this with respect to $\kappa$ numerically.  With a little effort one can see that in the thermodynamic limit, the PHF and UHF energies coincide.

Although the various expressions and matrix elements we have used can be derived using the angular momentum algebra discussed above or by using the fermionic anticommutation rules, mapping the problem onto a set of Schwinger bosons\cite{Lerma13} leads to particularly simple intermediate steps.

The idea behind the Schwinger boson mapping is to express the pseudo-spin operators $J_+$, $J_-$, and $J_z$ in terms of bosons rather than fermions.  Using $a^\dagger$ as the operator that puts a boson in the lower state and $b^\dagger$ as the operator that puts a boson in the upper state, we have
\begin{subequations}
\begin{align}
J_z &= \frac{1}{2} \, \left(b^\dagger \, b - a^\dagger \, a\right),
\\
J_+ &= b^\dagger \, a,
\\
J_- &= a^\dagger \, b.
\end{align}
\end{subequations}
The state with $n_a$ particles in the lower level and $n_b$ particles in the upper level is
\begin{equation}
|n_a,n_b\rangle = \frac{1}{\sqrt{n_a! \, n_b!}} \, \left(a^\dagger\right)^{n_a} \, \left(b^\dagger\right)^{n_b} |-\rangle_B,
\end{equation}
where $|-\rangle_B$ is the bosonic vacuum.  We require the wave function for an $n$-particle Lipkin Hamiltonian to have $n$ bosons, which guarantees the physicality of the mapping for the $j=n/2$ subspace.  The non-interacting ground state puts all bosons in the lower level, and in this language would be denoted as $|n,0\rangle$.  The states $|m\rangle$ defined above are in this bosonic language $|n-m,m\rangle$.

For the broken-symmetry mean-field, we have bosonic operators $\alpha^\dagger$ and $\beta^\dagger$ which put particles in the lower and upper broken-symmetry levels, and are given by
\begin{equation}
\begin{pmatrix}\alpha^\dagger \\ \beta^\dagger\end{pmatrix} = \frac{1}{\sqrt{1+\kappa^2}} \, \begin{pmatrix} 1 & \kappa \\ -\kappa & 1 \end{pmatrix} \, \begin{pmatrix} a^\dagger \\ b^\dagger \end{pmatrix}.
\end{equation}

\bibliography{LipkinBib}

\end{document}